\documentclass[showpacs,showkeys,aps,prl,twocolumn]{revtex4}
\usepackage{graphicx}
\usepackage{pstricks}

\newcommand{\degr}{\(^\circ\)}
\newcommand{\ignore}[1]{}
\newcommand{\etal}{{\it et al.}}

\begin{document}

\title{Ultrathin oxides: bulk-oxide-like model surfaces or unique films?}
\author{Christoph Freysoldt}
\author{Patrick Rinke}
\author{Matthias Scheffler}
\affiliation{Fritz-Haber-Institut der Max-Planck-Gesellschaft,
             Faradayweg 4--6, 14195 Berlin, Germany}

\begin{abstract}
To better understand the electronic and chemical properties of
wide-gap oxide surfaces at the atomic scale,
experimental work has focused on epitaxial films on metal substrates.
Recent findings show that these films are considerably thinner 
than previously thought. This raises doubts about the 
transferability of the results to surface properties of thicker films
and bulk crystals. 
By means of density-functional theory 
and approximate $GW$ corrections for the electronic spectra
 we demonstrate for 
three characteristic wide-gap oxides (silica, alumina, and hafnia)
the influence of the substrate and highlight critical differences between the 
ultrathin films and surfaces of bulk materials.
Our results imply that monolayer-thin oxide films have rather unique properties.
\end{abstract}


\maketitle

\begin{figure}
\includegraphics[width=0.47\textwidth]{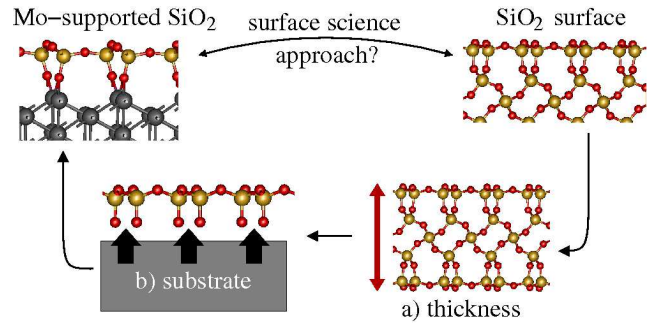}
\caption{
(Color online)
The critical assumptions in the surface science approach (here for SiO$_2$/Mo)
are tested by following a theoretical pathway that transforms
the surface of a bulk material (top right) to an epitaxial thin film 
on Mo(112) (top left):
a) decreasing the thickness and b) placing the film onto the substrate.
O atoms are depicted in red, Si in yellow and Mo in gray. }
\label{fig:surfsciappr}
\end{figure}

On the nanoscale materials often reveal new and unexpected features.
In particular ultrathin oxide films have
recently generated increasing interest.
First, thin oxide films themselves play an important role in technological
applications such as electronic devices, fuel cells, gas sensors,
corrosion and scratch protection, or heterogeneous catalysts.
Second, they are increasingly used as model oxides in the surface science
approach to thicker or bulk-like oxides:
Many experimental techniques
such as scanning tunneling microscopy or photoemission spectroscopy
cannot be directly applied to insulating materials, but 
require electrically conducting samples. Ultrathin epitaxial oxide
films on metal substrates overcome this limitation 
\cite{GoodmanModelCatalysts,FreundReview,
      Schroeder,Schroeder02,ChenSantraGoodman,
      Sauer,Pacchioni,GoodmanIsolated,Wendt,Jaeger,
      KresseAluminaNiAl,Kulawik}.
However, the direct transfer of experimental conclusions from these films to
bulk oxide surfaces relies on the two critical assumptions that
a) the thickness dependence of the film and b) substrate-film interactions
are negligible for the surface properties (cf. Fig.~\ref{fig:surfsciappr}).
In the present letter we argue that
the finite thickness and the presence of the substrate modify the film's
geometry and/or stoichiometry
in a way that is unknown and maybe even impossible
for surfaces of thicker films and bulk crystals.
In addition, the coupling of the film to the metal's Fermi level
may induce charge transfer \cite{PacchioniMgOPd}.
The resulting mechanical, electronic,
and chemical properties may then differ considerably
from the systems that were the original goal of the study.
Instead, these ultrathin films should be considered as new materials
with interesting and novel properties of their own.

For a meaningful comparison between thin films and bulk oxides
one has to first identify the oxide 
surfaces with the closest resemblance to the ultrathin film.
Then, one would like to isolate the influence
of the substrate from that of the finite thickness or the thickness
dependence in the surface science approach, to analyze each
contribution separately. 
While the latter is not possible with the
experimental tools available to date, it is easily achieved
in theory by the hypothetical two-step pathway shown in 
Fig.~\ref{fig:surfsciappr}: a) reduce the thickness of free-standing
films and study the effect of the thickness variation b) 
place an ultrathin film on a substrate analyzing the substrate dependence.
In this Letter, we employ density-functional theory
(DFT) in the local-density 
approximation (LDA) \cite{rem:RepeatedSlab} to address these questions
for three representative thin-film oxide systems: 
1) silica on Mo(112), a recent prototypical surface science
model for ceramic support materials of heterogeneous catalysts,
2) alumina, which in various modifications is also frequently applied as
model substrate for catalytic reactions,
and 3) hafnia, the material of choice in the most advanced microprocessor
technology \cite{ITRS}.

For silica, flat and well-or\-de\-red films have successfully been grown
on Mo(112) by \mbox{Freund} \etal{} and by Goodman \etal{}
\cite{Schroeder,Schroeder02,ChenSantraGoodman}. 
Recent \textit{ab initio} simulations \cite{Sauer,Pacchioni} 
suggest a two-dimensional network structure 
(isostructural, as we will show below, to the
$\alpha$-quartz (0001) surface \cite{Rignanese}),
whereas Goodman \etal{} favor an isolated [SiO$_4$] cluster model
\cite{GoodmanIsolated}.
Indications for thickness effects were given by 
Wendt \etal{}, who have demonstrated
in a comparative study including thicker (but amorphous) films 
that the electronic structure varies with the film
thickness \cite{Wendt}.
Alumina films have been grown on a wide
variety of substrates, e.g., Re, Mo, Ni$_3$Al, NiAl, or Fe$_3$Al.
A typical, intensively studied example is NiAl(110), where
a well-ordered alumina film of defined thickness is obtained
by a direct oxidation of the alloy's surface \cite{Jaeger}.
Its complex structure is governed by the interplay between the
local, preferentially hexagonal ordering of the oxide and
its linkage to the Al atoms at the interface to the 
metal alloy \cite{KresseAluminaNiAl}. 
Recent STM experiments for the adsorption of small gold cluster
on this film have shown that surface properties
are modified by the film's substrate \cite{Kulawik}.
Ultrathin hafnia films on metallic substrates
have not been investigated so far.

To investigate how far ultrathin films could reflect the 
surface properties of macroscopic samples \emph{in principle}  
we first exclude the substrate influence
by studying free-standing silica, alumina, and hafnia films. 
For all three oxides we find substantial differences between
monolayer (ML)-thin and thicker films.
We concentrate on the atomic structure
and the electronic density of states (DOS). Many other properties
are correlated to these two key aspects, e.g.,  the atomic configuration and
the local electronic structure determine the chemical activity.
The films are fully relaxed within the surface unit cell of the most
stable bulk surface, i.e.,
the (0001) surfaces of $\alpha$-quartz (silica) \cite{Rignanese}
and $\alpha$-alumina \cite{SchefflerAl2O3},
and the (111) surface of cubic hafnia \cite{rem:cubicHafnia}.
This ensures that, upon increasing the thickness, the films
systematically develop towards the bulk limit.

For silica, we have compared a variety of surface terminations and
reconstructions. We find that the most stable films are stoichiometric
and have fully saturated bonds. 
The structures represent perfect networks of [SiO$_4$]
units, in agreement with the trends in bulk silicates \cite{Liebau}. This 
similarity in the local geometry is contrasted by the flexibility of
the bonding network that easily adjusts to the constraints imposed
by the finite thickness and leads to quite different structures for
each thickness.
Films with more than 4 formula units 
are best described as $\alpha$-quartz-like layers sandwiched between
two reconstructed surfaces with three-membered silica rings
at the interface \cite{Rignanese} (Si$_5$O$_{10}$ being highly
distorted). In the Si$_4$O$_8$ film, which has no intermediate layers, 
the reconstructed surfaces are directly linked together.
Si$_2$O$_4$ consists of a single-layer network with two-membered silica rings,
a motif known from other silica surfaces \cite{Pacchioni,Ceresoli}.

\begin{figure}
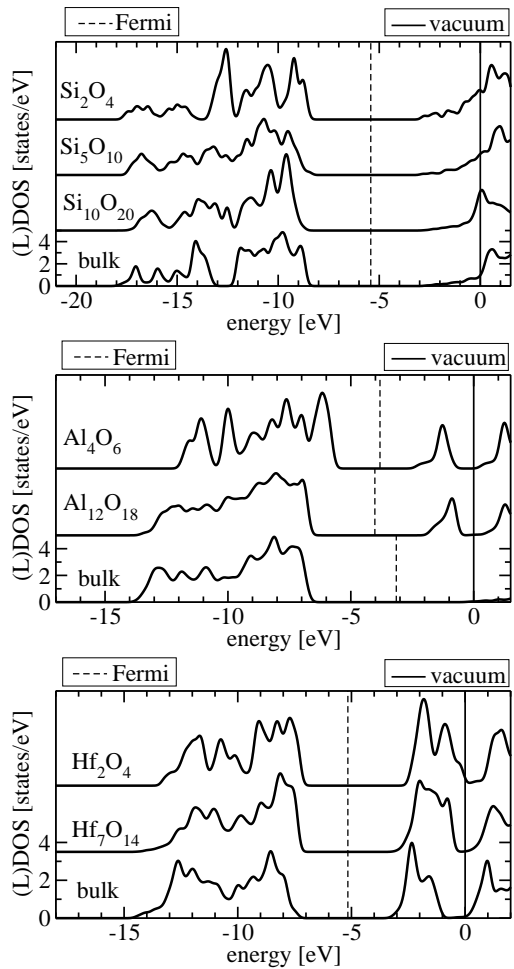

\includegraphics[scale=0.35,clip]{sio2-sdoscomp.eps}
\includegraphics[scale=0.35,clip]{al2o3-sdoscomp.eps}
\includegraphics[scale=0.35,clip]{hfo2-sdoscomp.eps}
\caption{Surface DOS for silica, alumina and hafnia films in comparison to
the bulk DOS.}
\label{fig:thickness}
\end{figure}

The structural variations in the alumina and hafnia films
are less drastic, but significant for ultrathin
films, too. For example, the outermost Al atom of the alumina surface
is known to strongly relax
inward \cite{SchefflerAl2O3}.
This relaxation shows a pronounced thickness dependence.
The interlayer spacing is reduced by 62\% for the Al$_4$O$_6$
film, by 96\% for Al$_6$O$_9$, and 87\% for Al$_8$O$_{12}$ and thicker
films.
The relaxations at the hafnia surface are
smaller ($<$4\% of the interlayer spacing),
but the thinnest films again deviate from their thicker counterparts.
The trend towards smaller structural changes reflects
the higher coordination in alumina (6/4 for Al/O,
respectively) and hafnia (8/4 for Hf/O) compared to silica (4/2 for Si/O) 
which limits the geometrical flexibility.

The effect of the finite thickness on the electronic and
chemical properties can be analyzed in terms of the DOS,
displayed for selected films in Fig.~\ref{fig:thickness}.
While the wide-gap insulating character of the perfect films is not 
affected by the thickness, the peak structure of the DOS differs significantly
for ultrathin films.
Since defect levels are most likely subject to
the same shifts as the bands from which they derive, this implies that 
ML-thin films may show critical deviations in their defect properties.
Fig.~\ref{fig:thickness} highlights
the differences in the local DOS at the surface
(the detailed evolution of the electronic structure 
will be discussed elsewhere) comparing 
the surface DOS of one or two ultrathin films to
a thick film, more representative for macroscopic samples, and the total
DOS of the bulk. In all three cases, 
the ultrathin films have a unique peak structure that differs from
that of thicker films. These varying peaks are not
associated with particular local orbitals but must be considered
properties of the overall electronic structure.
In addition, the Al$_4$O$_6$ film shows a
remarkable reduction of the band gap, while the valence band width
is comparable to that of thicker films and also the bulk. 
This indicates an overall weakening of the Madelung potential when
a substantial amount of the ions is undercoordinated \cite{PacchioniMgOPd}.
In all other cases, the band widths and gaps are close to
their bulk values. In general, we find that the changes in the
electronic structure follow those in the local atomic structure,
which underlines the crucial importance of the latter.

Of course, supported films are additionally
influenced by the interface to their substrate. Yet, our results for 
free-standing films
clearly demonstrate that the ML-thin films cannot be considered 
representative for the structural, electronic, defect, or surface 
properties of the respective bulk materials in contrast to earlier
statements \cite{GoodmanModelCatalysts,FreundReview,Schroeder,Jaeger}. 
Only when the films are thicker than 3-4 layers, the surface properties of
defect-free surfaces are approaching those of bulk-like
systems. This is a noticeable smaller film thickness than known
for metal surfaces, where for closed packed surfaces typically around 8 
in some cases even up to 20 layers \cite{AlSurface} are needed, 
but in line with other ionic materials such as NaCl \cite{BoLi}.
We attribute this trend to the degree of localization of the valence
electrons in these materials, which is larger in the insulators
(considered here) than in metals with typical semiconductors in between.
The deviating behavior of ML-thin films offers the prospect that
experimentally accessible ultrathin
oxide films such as silica on Mo(112) may be understood as new and unique
materials in their own right, opening possible new routes to 
devise insulator surfaces with novel properties.

We will now in more detail discuss the role of the substrate for
the example of silica on Mo(112), assuming the recently suggested
structure model \cite{Sauer,Pacchioni} (denoted as ``siloxane surface''): 
Corner-sharing [SiO$_4$] tetrahedrons form a two-dimensional,
perfectly bonded hexagonal network, that is 
linked to the Mo surface via oxygen (cf. Fig.~\ref{fig:surfsciappr}).
The film has a Si$_2$O$_5$ stoichiometry that does not naturally
occur in bulk silica.
A comparison to other known thin film and surface structures reveals
that it is identical to 
the ``dense reconstruction'' of the $\alpha$-quartz (0001) 
surface \cite{Rignanese},
the ``silicate adlayer'' on 4H and 6H SiC \cite{Bernhardt},
and the ``siloxane surface'' of clay minerals \cite{Sposito}.
The correspondence to the $\alpha$-quartz surface (which apparently has not
been recognized previously) makes the silica film on Mo(112) an ideal
candidate for studying the role of the metal substrate. For this,
we regard the $\alpha$-quartz layers below the siloxane surface
as an alternative, insulating substrate in place of Mo(112).

Focusing on the structural aspects first, we find that
the siloxane surface adapts to the substrate
by reorienting the [SiO$_4$]
tetrahedrons as rigid units, a behavior well-known for silicates \cite{Liebau}.
A substrate-independent, ``ideal'' siloxane surface structure
was determined from a free-standing siloxane double layer (Si$_4$O$_8$).
Its optimized lattice constant (5.24\,\AA) is very
close to the quasihexagonal c($2\!\times\!2$) Mo(112) 
superlattice (a=5.21\,\AA{} and $\alpha$=63\degr) and also to
the ($\sqrt3\!\times\!\sqrt3$)R30\degr{} pattern 
(5.25\,\AA) 
observed for SiC \cite{Bernhardt},
whereas the lattice mismatch to
$\alpha$-quartz (4.85\,\AA) amounts to 8\%. This provides a simple explanation
why the siloxane surface forms readily on the lattice-matched
substrates Mo(112) and 4H/6H-SiC (0001), while it has not been possible to
prepare it experimentally on $\alpha$-quartz.
A further important factor is the linkage of the silica film
to the Mo rows on Mo(112). The threefold
symmetry axis of the [SiO$_4$] tetrahedrons is tilted by
14\degr{} against the surface normal (cf. Fig.~\ref{fig:surfsciappr}).
This introduces a height difference (buckling) for the
surface atoms of $\sim$0.4\,\AA.
The siloxane surface on quartz, however, tilts only by 4\degr{} and 
buckles by 0.2\,\AA. 
We conclude that the siloxane surface on Mo(112), used in the surface 
science experiments, exhibits new features and deviates noticeable 
from the ideal $\alpha$-quartz surface despite the identical
chemical connectivity.
 
\begin{figure}
\includegraphics[scale=0.38,clip]{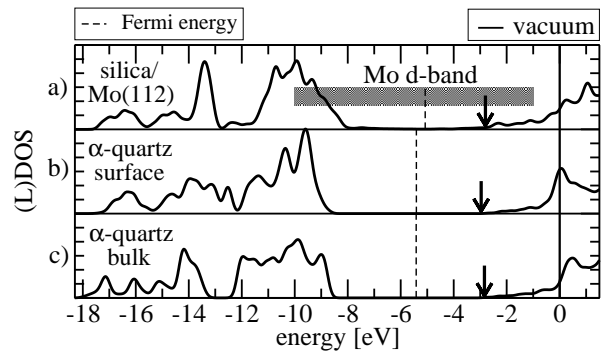}
\caption{Substrate influence on the silica surface DOS
for a) the Mo(112) and b) the $\alpha$-quartz substrates in comparison
to the total DOS of bulk $\alpha$-quartz c). 
The bottom of the surface conduction band is marked by an arrow.
The energetic position of the Mo $d$-bands is indicated by the gray bar.
}
\label{fig:substrate}
\end{figure}

To visualize the substrate's influence on the electronic structure,
the local DOS for the siloxane surface on Mo(112) and $\alpha$-quartz
are presented in Fig.~\ref{fig:substrate} along with the total DOS of
bulk $\alpha$-quartz as a reference.
The band edges at the surface agree 
with those of $\alpha$-quartz to within 0.5\,eV for both substrates.
There are no surface states in the large band gap. 
The Mo metal states, which energetically lie within the
surface band gap, play no important role at the surface.
However, the peak structure in the valence region of the surface DOS 
differs for the two substrates and deviates also from the DOS
of bulk $\alpha$-quartz. 
   We have analyzed the origin of these changes
by placing the silica/Mo surface structure on quartz
and find that mainly the structural variations are responsible
for the changes observed, whereas the chemically different interface
plays a minor role.
   
\begin{table}
\begin{ruledtabular}
\caption{Peak positions 
relative to the Fermi level for the electronic surface DOS of 
the siloxane surface on Mo(112) in comparison with experiment.
The energy scale is shifted by 5.1\,eV compared to
Fig.~\ref{fig:substrate}. The theoretical
data are not corrected for the estimated $GW$ shift of 2\,eV 
(see text).
}
\label{tab:peaks}
\begin{tabular}{l|ccccccc}
UPS$^a$    & --11.5 & --10.6    &  --    & --7.6  &  --   & --6.3 & --5.6$^c$ \\
MIES$^a$   & --11.6$^c$ & --10.4&  --    & --7.6  & --6.7 &  --   &  --   \\
\hline
DFT-LDA$^b$ & --9.5  & --8.3     & --7.3  & --5.6  & --4.9 & --4.3 & --3.8$^c$ \\
\end{tabular}

$^a$ From Fig.~4c) in Wendt \etal{} \cite{Wendt};
$^b$ this work; 
$^c$ shoulder
\end{ruledtabular}
\end{table}

That the surface electronic structure of the well-ordered silica
film on Mo(112) possesses a unique peak structure has been
observed in ultraviolet photoelectron spectroscopy (UPS) and
metastable impact electron spectroscopy (MIES) \cite{Schroeder02,Wendt}.
For comparing the DFT-LDA DOS to experimental spectra, we focus on
the peak \emph{positions} since the spectral \emph{intensities} depend on the
spectroscopic method and even make some peaks disappear in the
experimental MIE and UP spectra \cite{Wendt}.
To overcome limitations of DFT-KS in describing the quasiparticle spectrum
probed in experiment, we employ many-body perturbation theory 
in the $GW$ approximation \cite{Aulbur,Rinke}.
In $GW$, the electronic self-energy (which connects the non-interacting
electrons of DFT-KS to the truly interacting ones)
is given by the product of the
Green function $G$ and the screened interaction $W$.
$GW$ is typically applied as a correction scheme to the DFT-LDA energies
which corrects deficiencies of the LDA and includes quasiparticle
effects \cite{Aulbur,Rinke}.
We have performed $GW$ calculations for bulk $\alpha$-quartz.
Transferring in a first crude approximation the bulk corrections
to the silica film, we estimate that
the silica features in the valence region are shifted downward
by $\sim$2\,eV. 
Taking this $GW$ correction into account,  
the agreement with experiment (cf. Tab. \ref{tab:peaks}) becomes very good.
This agreement not only supports the siloxane surface as 
structural model for the silica film on Mo(112), but also
shows that the peaks observed in experiment are not generic
silica features, but specific to the actual film.

In light of these results we argue that the silica
film on Mo(112) must in fact be considered a novel oxidic surface, as
envisioned from free-standing films, and is not simply another incarnation
of a siloxane surface on bulk $\alpha$-quartz.
The lattice mismatch of 8\% to $\alpha$-quartz 
(which may prevent the formation of large ordered domains
in real samples) induces important changes in the structure compared 
to Mo(112), which subsequently modifies the electronic structure.

In summary we find that ultrathin films of wide-gap oxides require
two or three bulk-like layers to develop the characteristic
properties of the bulk materials and their surfaces. 
This implies that ML-films are no simple analog of thicker films or
realistic oxide surfaces. 
In addition, we demonstrated significant substrate-induced changes
in the surface
structure and electronic properties for the experimentally accessible silica
film on Mo(112).
We expect that similar conclusions hold for other metal-supported
oxide films.
Thus, ML-epitaxial thin films should not be seen merely as models for thick
oxides in heterogeneous catalysis or other applications. Instead they
may offer a unique way to devise entirely ``new'' insulator surfaces in
surface science or catalysis.


\begin{thebibliography}{28}
\bibitem[{rem({\natexlab{a}})}]{rem:RepeatedSlab}
\bibinfo{note}{We employ a plane-wave basis and norm-conserving
  pseudopotentials. The surfaces and thin films are modeled in the
  repeated-slab approach.}
\bibitem[{rem({\natexlab{b}})}]{rem:cubicHafnia}
\bibinfo{note}{We have chosen the cubic modification rather than the
  thermodynamically stable monoclinic one, because it appears to play an
  important role in hafnia thin films \cite{Aarik,hafniaMOCVD}.}
\end{thebibliography}

\end{document}